\documentclass[letterpaper]{article} 
\usepackage{aaai2026}  
\usepackage{times}  
\usepackage{helvet}  
\usepackage{courier}  
\usepackage[hyphens]{url}  
\usepackage{graphicx} 
\usepackage{natbib}  
\usepackage{caption} 
\usepackage{algorithm}
\usepackage{algorithmic}
\usepackage{multirow}
\usepackage{amsmath}
\usepackage{newfloat}
\usepackage{listings}
\DeclareCaptionStyle{ruled}{labelfont=normalfont,labelsep=colon,strut=off} 
\floatstyle{ruled}
\newfloat{listing}{tb}{lst}{}
\floatname{listing}{Listing}
\title{PSA-MF: Personality-Sentiment Aligned Multi-Level Fusion for Multimodal Sentiment Analysis}

\author{
    Heng Xie\textsuperscript{\rm 1}, Kang Zhu\textsuperscript{\rm 2}, Zhengqi Wen\textsuperscript{\rm 3}$^{*,}$, Jianhua Tao\textsuperscript{\rm 3,4}$^{*,}$, Xuefei Liu\textsuperscript{\rm 5}$^{*,}$, Ruibo Fu\textsuperscript{\rm 6}, Changsheng Li\textsuperscript{\rm 1}\thanks{Corresponding author}
}
\affiliations{
    \textsuperscript{\rm 1}School of Computer Science \& Technology, Beijing Institute of Technology, Beijing, China\\
    \textsuperscript{\rm 2}School of Computer Science, Wuhan University, Wuhan, China\\
    \textsuperscript{\rm 3}Beijing National Research Center for Information Science and Technology, Tsinghua University, Beijing, China\\
    \textsuperscript{\rm 4}Department of Automation, Tsinghua University, Beijing, China\\
    \textsuperscript{\rm 5}College of Computer and Information Engineering, Tianjin Normal University, Tianjin, China\\
    \textsuperscript{\rm 6}Institute of Automation, Chinese Academy of Sciences, Beijing, China
}

\usepackage{bibentry}

\begin{document}

\maketitle

\begin{abstract}
Multimodal sentiment analysis (MSA) is a research field that recognizes human sentiments by combining textual, visual, and audio modalities. The main challenge lies in integrating sentiment-related information from different modalities, which typically arises during the unimodal feature extraction phase and the multimodal feature fusion phase. Existing methods extract only shallow information from unimodal features during the extraction phase, neglecting sentimental differences across different personalities. During the fusion phase, they directly merge the feature information from each modality without considering differences at the feature level. This ultimately affects the model's recognition performance. To address this problem, we propose a personality-sentiment aligned multi-level fusion framework. We introduce personality traits during the feature extraction phase and propose a novel personality-sentiment alignment method to obtain personalized sentiment embeddings from the textual modality for the first time. In the fusion phase, we introduce a novel multi-level fusion method. This method gradually integrates sentimental information from textual, visual, and audio modalities through multimodal pre-fusion and a multi-level enhanced fusion strategy. Our method has been evaluated through multiple experiments on two commonly used datasets, achieving state-of-the-art results.
\end{abstract}


\section{Introduction}

With the rapid development of the internet and social media, user-generated video data has increased significantly. To enhance user experience, research on multimodal sentiment analysis (MSA) has attracted growing attention and expanded to applications in human-computer interaction \cite{cambria2017affective}, risk prediction \cite{zhu2023multimodal}, psychotherapy \cite{yang2018integrating}, and other fields \cite{das2023multimodal, fan2024icaps}. Compared to unimodal sentiment analysis, MSA integrates data from multiple modalities, such as text, visual, and audio, into a unified sentiment representation. This modal complementarity is more conducive to accurate and stable sentiment recognition \cite{gandhi2023multimodal, han2021improving}. However, it also presents two key challenges: how to extract effective sentimental representations from each modality, and how to effectively combine sentiment information from multiple modalities.

Existing methods approach the problem from the perspectives of unimodal feature extraction and multimodal fusion. Initially, various unimodal features are extracted using pre-trained models or other unimodal feature extraction tools \cite{caschera2016sentiment, xie2023multimodal, yang2022multimodal}. On this basis, multimodal semantic-level fusion is further performed. From the feature extraction perspective, methods include unimodal feature encoding \cite{sun2022learning}, modality feature decomposition \cite{yang2023confede}, and contrastive learning \cite{lin2022modeling}. From the fusion perspective, methods include tensor-based fusion \cite{liu2018efficient}, attention mechanism-based fusion \cite{wei2020multi}, and pre-trained model-based fusion \cite{rahman2020integrating}. However, current methods have issues: features extracted in the encoding phase do not incorporate personalized sentimental information and fail to account for differences in sentimental expression across personalities \cite{zhang2019persemon}. These personality differences significantly affect sentimental expression and perception in real-world scenarios \cite{zhu2024transferring}. Additionally, due to the heterogeneity of different modal data, extracting unimodal information for basic fusion cannot fully understand complex sentimental states \cite{han2021bi}. Traditional fusion strategies struggle to achieve deep interactions between modalities.

To address these problems, we propose a personality-sentiment aligned multi-level fusion (PSA-MF) method for MSA. From the feature extraction perspective, we integrate personality information into sentiment features by combining a pre-trained personality model with fine-tuned BERT to extract personalized sentiment features from text. 
We then achieve alignment between sentiment and personality through contrastive learning and a personalized sentimental constraint loss. From the perspective of fusion, we design a multi-level fusion method. First, the deep layers of BERT are used as the multimodal pre-fusion layer, combining shallow text embeddings with visual and audio features for preliminary alignment between modalities. Second, the output from the multimodal pre-fusion layer serves as a query for cross-modal interaction with visual and audio modalities. This process guides the personalized generation of visual and audio features while reducing information bias from modality distribution differences. Finally, through serial and parallel fusion, semantic-level fusion of personalized sentiments is conducted from multiple angles, enhancing cross-modal transmission of personalized sentimental cues while maintaining the complementary characteristics of each modality. This refined fusion architecture overcomes modality noise interference in early fusion and avoids the shortcomings of late fusion that neglect cross-modal correlations, achieving a balance between global consistency and local specificity in sentimental semantics.

Our contributions are as follows.

\begin{itemize}
\item We are the first to combine personality pre-trained models with a phase-wise fine-tuned BERT architecture to extract personalized sentimental features from the textual modality. This addresses the limitation of traditional methods that ignore personality factors.

\item We propose an innovative personality-sentiment alignment strategy that uses contrastive learning to achieve this alignment at the semantic level. Furthermore, we construct a constraint function that dynamically adjusts the alignment strength while confining the alignment process within the appropriate sentiment space.

\item We propose a multi-level fusion architecture that includes pre-fusion, cross-modal interaction, and enhanced fusion. The pre-fusion features direct audio-visual modalities toward personalized feature generation in the cross-modal module. Finally, by combining serial fusion and parallel fusion in a collaborative mechanism, we effectively balance global sentimental consistency with local modality complementarity.
\end{itemize}
\section{Related Work}

\subsection{Personality-Sentiment Interplay}

In the current field of affective computing, personality and sentiment analysis are considered two core elements. 
Eladhari et al. \cite{eladhari2014mind} proposed a personality-sentiment mapping model that maps different sentiments to corresponding personality traits. They developed a mental module applied in gaming, which provides a foundation for understanding the dynamic relationship between sentiments and personality.
To conduct in-depth research on personality recognition, Ponce-L{\'o}pez et al. \cite{ponce2016chalearn} created a Big Five personality dataset by screening 10,000 YouTube videos through fitting a Bradley-Terry-Luce model with maximum likelihood. This dataset has facilitated further research on the Big Five personality traits and sentiment analysis.
Since most studies on sentiment and personality analysis are conducted separately, Zhang et al. \cite{zhang2019persemon} built upon Ponce-L{\'o}pez's work to combine personality and sentiment for analysis. They constructed an end-to-end network structure and demonstrated the interrelationship between sentiment and personality traits through a multi-task learning strategy.
Mohammadi et al. \cite{mohammadi2020multi} used a component-based framework to analyze sentimental experiences and explored the relationship between personality traits and sentiments. The study found that the component model could effectively predict individual differences in discrete sentiments. It also emphasized the important role of personality in the generation of sentiments.
With the rapid iteration of generative technologies, dialogue systems based on large language models have also matured. Ma et al. \cite{ma2020survey} further explored dialogue systems that are more humanlike, incorporating sentimental awareness and personality awareness as key features of the system.

The above research demonstrates that it is crucial to consider the influence of personality traits in sentiment analysis. They provide a theoretical foundation and experimental support for understanding the complex relationship between sentiments and personality traits. Based on this, we are applying a personality pre-trained model in a sentiment analysis framework and design a personality-sentiment alignment module. This method effectively addresses the limitation of traditional approaches that ignore personality factors and restrict sentimental recognition performance. It also solves the problem that existing models cannot conduct personalized sentiment research on datasets with only sentiment labels.

\subsection{Multimodal Fusion Paradigms}
Existing multimodal works typically utilize unimodal encoders to extract features and employ multimodal encoders for fusion, combining encoders with varying parameter counts for tasks such as classification or retrieval. In early multimodal work, visual semantic embedding (VSE) \cite{chen2021learning} proposed that the visual encoder is heavier compared to the text embedding, and multimodal interactions are relatively simple. Building on this, CLIP \cite{radford2021learning} elevated the text encoder to the same scale as the visual encoder, and the multimodal aspect used contrastive learning to achieve the alignment of visual and textual semantic information, making the fusion part lightweight.
The work of Lu and Chen et al. \cite{lu2019vilbert, chen2020uniter} focused on modal interaction, achieving significant performance improvements by replacing traditional fusion methods such as concatenation and dot product with transformer blocks. To improve inference speed, Kim et al. \cite{kim2021vilt} drew inspiration from ViT \cite{dosovitskiy2020image}, replacing the visual encoder with patch embedding and using lightweight text and visual encoders for feature extraction, while employing heavier transformer blocks for multimodal fusion. To balance model performance and speed, Li et al. \cite{li2021align} use transformer-based pre-trained models as unimodal encoders and utilize the deeper layers of the text encoder as multimodal encoders.

In multimodal research, sentiment analysis is an emerging direction that typically combines information from text, visual, and audio modalities for sentiment classification. Early research adopted early fusion \cite{morency2011towards} and late fusion \cite{nojavanasghari2016deep} methods.
Current research utilizes pre-trained models to extract features from the textual modality, while visual and audio features are usually pre-extracted. Therefore, researchers are more focused on effectively fusing sentiment information from different modalities. This can be specifically divided into two aspects: decoupled feature fusion and direct fusion. 

Decoupling relies on sophisticated constraint conditions and requires the design of multiple loss functions, making the model relatively complex. For example, Zeng et al. \cite{zeng2024disentanglement} proposed a DTN that employs multiple encoders and decoders to map features from various modalities into homogeneous and heterogeneous spaces. Constraints are applied to the heterogeneous features output by the decoder to distinguish between homogeneous and heterogeneous features. Finally, homogeneous features are fused for sentiment classification. Direct fusion, on the other hand, focuses more on the multimodal fusion framework. For example, Yang et al. \cite{yang2020cm} first utilized BERT and COVAREP \cite{degottex2014covarep} to extract feature information from text and audio modalities, respectively. To achieve word-level modality alignment, they padded the audio sequence with zero vectors and finally employed masked multimodal attention for fusion.

\begin{figure*}[t]
\centering
\includegraphics[width=0.88\textwidth]{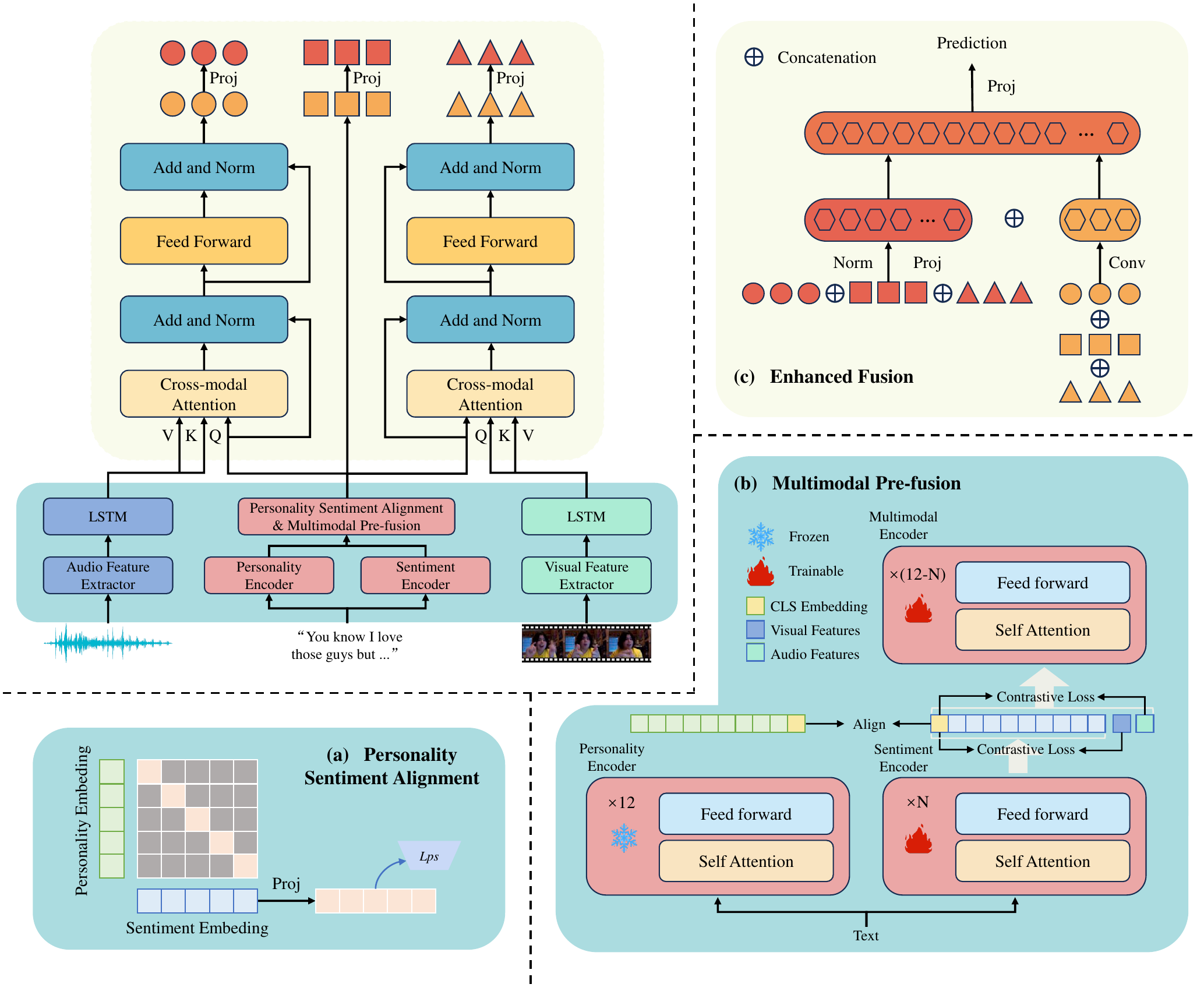}\\
\caption{The proposed personality-sentiment aligned multi-level fusion model (PSA-MF): In the top left corner of the diagram is the main framework of the model, which includes feature extraction and cross-attention modal interaction. (a) shows the personality-sentiment alignment module, which includes personality-sentiment contrastive learning and personalized sentimental constraints. (b) displays the multimodal pre-fusion module of the model, which utilizes the deep layers of BERT as a multimodal encoder for initial alignment across three modalities. (c) depicts the enhanced fusion module, which performs serial fusion and parallel fusion of the multi-level features from the upper-layer cross-attention, culminating in the final prediction.}
\label{fig1}
\end{figure*}

In current MSA, pre-extracted features are commonly used for audio and video, while text features are extracted through pre-trained models. This results in text representation spaces having significantly higher information density than visual and audio modalities. Simple same-level feature fusion makes it difficult to coordinate information density and semantic depth between different modalities. Therefore, we constructed a multi-level fusion structure that gradually alleviates the semantic gap between modalities through progressive feature interaction.

\section{Methodology}
In this section, we provide a detailed description of our method, with the model architecture shown in Figure \ref{fig1}.

\subsection{Unimodal Feature Extraction}
In the feature extraction phase, we utilize pre-trained models to extract sentiment and personality features from the textual modality, and apply LSTM models to extract visual and audio features.

\textbf{Textual Sentiment and Personality Feature Extraction.} We use two pre-trained models, the first N layers of a fine-tuned ${\text{BERT}_s}^1$ and Personality ${\text{BERT}p}^2$ to extract sentiment and personality embeddings from text, respectively. Given the text input \( X_t \), the representation for both sentimental and personality features can be expressed as:

\begin{equation}
\text{CLS}_i = \text{BERT}_i(X_t; \theta_{\text{BERT}_i}), \quad i \in \{s,p\}
\end{equation}

$\text{CLS}_i$ represents the output embeddings of the [CLS] token from the BERT model, which captures different types of features based on the value of $i$. $\theta_{\text{BERT}_i}$ denotes the parameters of the respective BERT models. Here, $i \in \{s,p\}$ where $s$ indicates the sentiment and $p$ indicates the personality.

\footnotetext[1]{https://huggingface.co/google-bert/bert-base-uncased}
\footnotetext[2]{https://huggingface.co/Minej/bert-base-personality}

\textbf{Visual and Audio Feature Extraction.} We use LSTM to process visual and audio features. Given the visual input \( X_v \) and audio input \( X_a \), the LSTM encoding outputs are:

\begin{equation}
h_m = \text{LSTM}(X_m; \theta_{\mathrm{LSTM}_m}), \quad m \in \{v,a\}
\end{equation}

where \(h_m\) represents the temporal feature embeddings extracted from different modalities, and \(\theta_{\mathrm{LSTM_m}}\) denotes the parameters of the LSTM models, with \(m \in \{v, a\}\) indicating visual and audio modalities, respectively.

\subsection{Personality-sentiment Alignment}

To align sentimental and personality features, we introduce contrastive learning. We first map these features to a common lower-dimensional space using linear transformations.

\begin{equation}
T_i = W_i \cdot \text{CLS}_i, \quad i \in \{s,p\}
\end{equation}

where \(W_i\) is a linear learning mapping matrix and \(T_i\) represents the textual mapped embeddings, with \(i \in \{s,p\}\) indicating sentiment and personality, respectively.

Specifically, the contrastive learning loss $\mathcal{L}_{cl}$ follows the CLIP \cite{radford2021learning} formulation:

\begin{equation}
\mathcal{L}_{cl} = -\log \frac{\exp(\text{sim}({T_s}^i, {T_p}^i)/\tau)}{\sum_{j=1}^N \exp(\text{sim}({T_s}^i, {T_p}^j)/\tau)}
\end{equation}

where the total number of samples is \( N \) with \( i,j \in [1,N] \) denoting sample indices, \(\text{sim}(\cdot, \cdot)\) denotes the computation of cosine similarity, and \( \tau \) is a temperature parameter that controls the smoothness of the feature distribution. This loss function encourages matched sentiment-personality pairs (positive pairs) to be closer in the feature space while pushing unmatched pairs (negative pairs) apart. Combined with cosine similarity, we define the compound contrastive learning loss:

\begin{equation}
\mathcal{L}_{ccl} =  \text{sim}({T_s}^i, {T_p}^i) \cdot \mathcal{L}_{cl} 
\end{equation}

where \(i\) denotes the \(i\)-th sample.

To further enhance the alignment of personalized sentiment features, we design a personalized sentimental constraint loss. We calculate the regression loss of personalized sentiment features using true sentimental labels \(y_i\):

\begin{equation}
\mathcal{L}_{\text{ps}} = (1 - \text{sim}({T_s}^i, {T_p}^i)) \cdot \| W_y \cdot {T_s}^i - y_i \|_1
\end{equation}

where \(W_y\) is a linear mapping and \(\| \cdot \|_1\) denotes the L1 norm. This step optimizes the model's error in personalized sentimental modeling tasks while balancing the integration of sentiment features with personality.

The overall personality-sentiment alignment loss is a combination of the two losses:

\begin{equation}
\mathcal{L}_{Align} = \mathcal{L}_{ccl} + \mathcal{L}_{ps}
\end{equation}

In summary, contrastive learning aligns sentiment with personality, and the personalized sentimental constraint loss further refines this alignment by constraining the difference between personalized features and true labels, effectively enhancing the model's sentiment classification performance.

\textbf{Multimodal Pre-fusion Module.}  
The feature information of each modality has been extracted through the above modules, including the personality information features from the textual modality. We begin the multimodal pre-fusion of the features above. First, we perform cross-modal contrastive learning between the textual modality and the other two modalities, with the loss function designed as follows:

\begin{equation}
\mathcal{L}_{clm} = \sum_{m \in \{v, a\}} -\log \frac{\exp\left(\text{sim}({\text{CLS}_s^i}, {X_m^i})/\tau\right)}{\sum_{j=1}^N \exp\left(\text{sim}({\text{CLS}_s^i}, X_m^j)/\tau\right)}
\end{equation}

where \( \text{CLS}_s^i \) represents the [CLS] token of the text for the \(i\)-th sample, while \( X_v^i \) and \( X_a^i \) correspond to the visual and audio features of the same sample, respectively. \(\mathcal{L}_{clm}\) is the total contrastive learning loss for the text and both modalities \( v \) and \( a \).

Then, we use the last (12-N) layers of the fine-tuned BERT as a multimodal encoder. The features from the three modalities are concatenated and input for pre-fusion. Through bidirectional encoding, contextual information is effectively captured, particularly enhancing the semantic association capture in cross-modal space and the progressive transfer of personality representation. This produces the multimodal pre-fusion embedding.

\begin{equation}
\text{CLS}_m = \text{BERT}_m([CLS_s, X_v, X_a]; \theta_{\text{BERT}_m})
\end{equation}

where, \([CLS_s, X_v, X_a]\) represents the concatenated feature vectors of the three modalities, and \(\theta_{\text{BERT}_m}\) denotes the parameters of the multimodal encoder, resulting in the multimodal pre-fusion embedding \(\text{CLS}_m\).

\textbf{Cross-modal Interaction Module.} The cross-modal personality representation output from the pre-fusion stage interacts with the visual and audio modalities. In this process, the pre-fusion features serve as query vectors for attention, guiding personalized weight allocation for the other two modalities, while further eliminating stepwise shifts between heterogeneous modalities, achieving personality-driven modality-specific reconstruction. The feature representation that fuses visual and audio is:

\begin{equation}
V_t = \text{Att}_v(M_s, h_v, h_v)
\end{equation}

\begin{equation}
A_t = \text{Att}_a(M_s, h_a, h_a)
\end{equation}

where \(M_s\) represents the pre-fusion features after linear transformation of \(\text{CLS}_m\), and \(h_v\) and \(h_a\) are the temporal features of the visual and audio modalities, respectively. Additionally, \(\text{Att}_v\) and \(\text{Att}_a\) are multi-head attention mechanisms. The fusion features \(V_t\), \(A_t\), and \(M_s\) undergo linear transformations to obtain the cross-modal sentimental enhancement representations \(V'_t\), \(A'_t\), and \(M'_s\).

\textbf{Enhanced Fusion Module.} We perform enhanced fusion on the features obtained from the upper-level cross-modal interaction. This is aimed at strengthening the propagation of personality and sentimental information across modalities and enhancing the complementarity of information within each modality. Specifically, we first construct a dual-stream network with serial fusion and parallel fusion. Then, we concatenate the fusion results and pass them through a subnet for the final prediction.

The serial fusion layer uses a linear layer to merge the feature dimensions of the cross-modal sentimental enhancement representations \(V'_t\), \(A'_t\), and \(M'_s\):

\begin{equation}
F_s = W_p \cdot [V'_t, A'_t, M'_s]
\end{equation}

where \(W_p \) represents the learnable weight matrix, and \([V'_t, A'_t, M'_s]\) denotes the concatenated feature vectors. The LayerNorm operation ensures stable training by normalizing the features across the batch dimension. The fully connected layer further facilitates the transfer of personalized sentiments between different modalities while allowing for complementary learning among the various modalities.

The parallel fusion layer uses convolution operations to merge the stacked fusion features \(V_t\), \(A_t\), and \(M_s\):

\begin{equation}
F_p = \text{Conv}([V_t, A_t, M_s])
\end{equation}

where \text{Conv} represents a convolutional operation with kernel size 3, which transforms the concatenated features \([V_t, A_t, M_s]\) into a compressed representation, effectively capturing local spatial information across modalities.

The enhanced fusion features can be obtained by connecting two fusion paths. These enhanced fusion features are then fed into the final subnet to obtain the prediction result:

\begin{equation}
\hat y = Sub([F_s, F_p])
\end{equation}

where \(\hat y\) represents the final prediction result of the model, and \([F_s, F_p]\) denotes the connected enhanced fusion features. The \text{Sub} is a fully connected layer network structure with two layers of non-linear activation. The final sentiment prediction is made by enhancing and fusing personalized sentiment features from multiple perspectives.

The total loss function includes the loss of personality-sentiment alignment, the loss of cross-modal contrastive learning, and the loss L1 between predicted results and ground truth. The total loss function is expressed as:
\begin{equation}
\mathcal{L}_{Total} = \mathcal{L}_{Align} + \mathcal{L}_{clm} + \| \hat y_i - y_i \|_1
\end{equation}

\begin{table*}[h]
\centering
\caption{Comparing the results of various baseline methods on the CMU-MOSI and CMU-MOSEI datasets.}
\label{tab:exper}
\resizebox{\textwidth}{!}{%
\begin{tabular}{*{11}{c}}
\hline
\multirow{2}{*}{\textbf{Method}} & \multicolumn{5}{c}{\textbf{MOSI}} & \multicolumn{5}{c}{\textbf{MOSEI}} \\
\cline{2-11}
 & \makebox[0.03\textwidth][c]{\textbf{MAE}$\downarrow$} & \makebox[0.03\textwidth][c]{\textbf{Corr}$\uparrow$} & \makebox[0.02\textwidth][c]{\textbf{Acc7}$\uparrow$} & \makebox[0.03\textwidth][c]{\textbf{Acc2}$\uparrow$} & \makebox[0.03\textwidth][c]{\textbf{F1}$\uparrow$} & \makebox[0.03\textwidth][c]{\textbf{MAE}$\downarrow$} & \makebox[0.03\textwidth][c]{\textbf{Corr}$\uparrow$} & \makebox[0.02\textwidth][c]{\textbf{Acc7}$\uparrow$} & \makebox[0.03\textwidth][c]{\textbf{Acc2}$\uparrow$} & \makebox[0.03\textwidth][c]{\textbf{F1}$\uparrow$} \\
\hline
TFN & 0.901 & 0.698 & 34.9 & -/80.8 & -/80.7 & 0.593 & 0.700 & 50.2 & -/82.5 & -/82.1 \\
LMF & 0.914 & 0.689 & 33.2 & 79.8/- & 79.5/- & 0.680 & 0.590 & 48.0 & 80.4/- & 78.2/- \\
MuLT & 0.871 & 0.698 & - & -/83.0 & -/82.8 & 0.580 & 0.703 & - & -/82.5 & -/82.3 \\
MISA & 0.783 & 0.761 & 42.3 & 81.8/83.4 & 81.7/83.6 & 0.555 & 0.756 & - & 83.6/85.5 & 83.8/85.3 \\

MVCL & 0.769 & 0.783 & - & 83.7 & 84.2 & 0.573 & 0.741 & - & 85.0 & 85.0 \\
HyCon & 0.713 & 0.790 & 46.6 & 85.2 & 85.1 & 0.601 & \textbf{0.776} & 52.8 & 85.4 & 85.1 \\
PriSA & 0.714 & {0.792} & 47.3 & {83.38/85.52} & {83.24/85.45} & {0.523} & {0.772} & 54.65 & {82.84/85.93} & {83.18/85.87} \\
FGTI & {0.702} & {0.791} & \textbf{48.0} & {85.8} & {85.8} & {0.536} & {0.771} & {53.4} & {86.0} & {86.0} \\
ULMD & {0.700} & {0.799} & {47.8} & {85.82} & {85.71} & {0.531} & {0.770} & {53.81} & {85.95} & {85.91} \\
\hline
\textbf{PSA-MF (Ours)} & \textbf{0.686} & \textbf{0.807} & 46.5 & \textbf{83.67/86.43} & \textbf{83.30/86.19} & \textbf{0.521} & 0.774 & \textbf{55.0} & \textbf{83.76/86.30} & \textbf{84.21/86.28} \\
\hline
\end{tabular}%
}
\end{table*}

\section{Experiments}
\subsection{Datasets}
We used two classic MSA datasets to validate the effectiveness of our model.

\textbf{CMU-MOSI.} The CMU-MOSI dataset is the first opinion-level annotated corpus for sentiment and subjectivity analysis of online videos. This dataset collects movie review videos from YouTube and contains 2,199 opinion segments. The sentiment annotations for each sample range from -3 (highly negative) to 3 (highly positive).

\textbf{CMU-MOSEI.} The CMU-MOSEI dataset is the next generation of CMU-MOSI, designed to collect a broader variety of samples from different speakers and topics. This dataset includes 23,453 annotated video clips from YouTube, with sentiment annotations similar to those in CMU-MOSI.

\subsection{Experimental Setup}
Our method is implemented with PyTorch and trained on an NVIDIA RTX 3090 GPU. Initial learning rates for MOSI and MOSEI datasets are \(10^{-4}\), with batch sizes of 128 and 256, respectively. For consistency with existing work, video embeddings derive from 35 facial action units extracted using Facet \cite{littlewort2011computer}, while audio embeddings come from features extracted using COVAREP \cite{degottex2014covarep}. We evaluated performance on regression and classification tasks using mean absolute error (MAE), Pearson correlation coefficient (Corr), binary classification accuracy (Acc2), seven-class classification accuracy (Acc7), and F1 score (F1).

\subsection{Baselines}
We compared our method with several state-of-the-art approaches in MSA, including both classic and recent advanced methods:

Tensor Fusion Methods: TFN \cite{zadeh2017tensor} and LMF \cite{liu2018efficient} achieve multimodal tensor-level fusion through specialized tensor fusion layers.

Cross-modal Attention and Encoding: MuLT \cite{tsai2019multimodal} and MISA \cite{hazarika2020misa} utilize cross-modal attention mechanisms and specific modality feature encoding methods, respectively.

Contrastive Learning: MVCL \cite{liu2023improving} and HyCon \cite{mai2022hybrid} employ various contrastive learning techniques to learn modality differences and extract effective sentimental representations.

Priority-based Fusion: PriSA \cite{ma2023multimodal} utilizes priority-based fusion to guide inter-modal correlation computation and employs distance-aware contrastive learning to handle mixed-modal correlations.

Multi-granularity Fusion: FGTI \cite{zhi2024fine} and MMIN \cite{fang2024multi} focus on reducing modality redundancy and enhancing modality-invariant representations through multi-granularity fusion approaches.

Feature Decoupling Method: ULMD \cite{zhu2025multimodal} achieves feature decoupling by introducing modality separators and a unimodal label generation module, dividing representations into invariant and specific components.

\subsection{Experimental Results}
Table \ref{tab:exper} presents the comprehensive experimental results of our proposed method compared to the baseline approaches on the MOSI and MOSEI datasets. Our method consistently outperforms existing state-of-the-art models across almost all metrics, demonstrating its effectiveness in MSA.

Compared to traditional tensor fusion methods like TFN and LMF, our approach shows significant improvements. On the MOSI dataset, our method achieves a 4.8\% increase in Acc2 and a 4.6\% increase in F1 compared to TFN. This substantial improvement can be attributed to our incorporation of personality information at the feature extraction level, which enables the model to capture individual differences in sentimental expression more accurately.

Compared to methods that utilize cross-modal attention mechanisms and specific modal encoding (MuLT and MISA), our approach still shows advantages. On the MOSEI dataset, our method improved Acc2 by 3.2\% and F1 by 3\% compared to MISA. This enhancement stems from the multi-level fusion mechanism of our model. During the multimodal pre-fusion stage and the cross-modal interaction stage, the model achieves a dynamic transfer of personalized sentimental features across modalities, initially bridging the modal differences. In the enhanced fusion stage, the model further strengthens these capabilities.

Our method also outperforms advanced techniques that utilize contrastive learning (MVCL and HyCon). This can be attributed to our pioneering introduction of personality-sentiment contrastive learning in addition to cross-modal contrastive learning. Compared to their methods, we simultaneously consider both multimodal semantic differences and personalized sentimental differences.

Compared to state-of-the-art methods, our approach achieves 0.91\% improvement in Acc2 and 0.74\% increase in F1 over PriSA on the MOSI dataset, and 0.63\% improvement in Acc2 and 0.39\% increase in F1 over FGTI. Unlike PriSA, which uses priority-based fusion for inter-modal correlation and distance-aware contrastive learning, our personality-sentiment alignment and multi-level fusion better capture both fine-grained and high-level sentimental cues. While PriSA handles mixed-modal correlations, our method uses personality features for more nuanced, personalized sentiment analysis. FGTI focuses on multi-granularity fusion and reduces modal gaps through similarity metrics and self-supervised learning. Although FGTI enhances modal specificity through constraints, our method uses personality information for more differentiated sentiment understanding across modalities.

Compared to decoupling methods, our approach still maintains excellent performance. On the MOSI dataset, our method improved Acc2 and F1 by 0.61\% and 0.48\%, respectively. In contrast to ULMD, which decouples modalities and features separately, our approach is simpler. It achieves the alignment of personality and sentiment solely by introducing constraint functions. By improving BERT to construct a multimodal encoder and combining multi-level fusion, we effectively alleviate the heterogeneity of multimodal data. There is no need for an additional decoupling network, which avoids excessive constraint functions and is beneficial for model training and convergence.

Compared to existing advanced models in different datasets, our model leads in almost all metrics. These experimental results demonstrate the efficacy of our personality-sentiment alignment method. By progressively integrating multimodal information and more comprehensively learning the complex relationships between sentiments and personality, our model demonstrates a stronger ability to capture subtle sentimental expressions and improve overall sentiment analysis accuracy.

\subsection{Ablation Study}
To validate each component in our model, we conducted a comprehensive ablation study on the MOSI dataset, which mainly includes module ablation and loss function ablation. Results are shown in Table \ref{tab:module ablation}.

\begin{table}[h]
\centering
\caption{Module and loss function ablation study on the CMU-MOSI.}
\resizebox{0.48\textwidth}{!}{ 
\label{tab:module ablation}
\begin{tabular}{l c c c c c}
\hline
\textbf{Model} & \textbf{MAE$\downarrow$} & \textbf{Corr$\uparrow$} & \textbf{Acc-7$\uparrow$} & \textbf{Acc-2$\uparrow$} & \textbf{F1$\uparrow$} \\
\hline
w/o-PF & 0.711 & 0.795 & 44.6 & 84.60 & 84.47 \\
w/o-BF & 0.735 & 0.778 & 43.2 & 84.76 & 84.44 \\
w/o-EF & 0.806 & 0.788 & 37.3 & 85.21 & 84.96 \\
w/o-($\mathcal{L}_{\text{ps}}$) & 0.724 & 0.784 & 44.5 & 83.99 & 83.92 \\
w/o-($\mathcal{L}_{\text{clm}}$) & 0.754 & 0.784 & 40.8 & 85.06 & 84.92 \\
\textbf{PSA-MF(Ours)} & \textbf{0.686} & \textbf{0.807} & \textbf{46.5} & \textbf{86.43} & \textbf{86.19} \\
\hline
\end{tabular}
}
\end{table}

\textbf{Module Ablation Study.} We investigated removing key components: the personality feature extraction (w/o-PF), the BERT-based multimodal pre-fusion (w/o-BF), and the enhanced fusion module (w/o-MF).

Removing the personality feature extraction caused the most significant decline, with Acc2 dropping by 1.83\% and F1 by 1.72\%. This highlights the critical role of personality information in obtaining personalized sentimental features. By incorporating personality information, the model recognizes subtle differences in sentimental expression among individuals, improving sentiment recognition accuracy.

Removing the BERT-based multimodal pre-fusion decreased performance, with Acc2 dropping by 1.67\% and F1 by 1.75\%. This indicates its crucial role in alleviating modal heterogeneity and facilitating the dynamic transfer of personality information across modalities in early fusion.

Removing the multi-level fusion decreased performance, with Acc2 dropping by 1.22\% and F1 by 1.23\%. This demonstrates enhanced fusion strengthens cross-modal transfer of personalized sentimental information while maintaining complementarity across modalities. It enables capturing fine-grained and high-level sentimental cues, improving sentiment classification capability.

\textbf{Loss Function Ablation Study.} We also investigated the impact of removing loss functions: the personalized sentiment constraint loss (w/o-($\mathcal{L}_{\text{ps}}$)) and the multimodal contrastive learning loss (w/o-($\mathcal{L}_{\text{clm}}$)).

Removing the personalized sentimental constraint loss caused the most significant decline among all ablation experiments, with Acc2 dropping by 2.44\% and F1 by 2.27\%. This indicates that this loss plays a crucial role in balancing during training. It guides the model to align personality traits with appropriate sentimental expressions while ensuring sentiment classification accuracy by dynamically adjusting the strength of personality-sentiment alignment. Specifically, the constraint loss limits the deviation between the personalized sentimental representation and the sentiment labels, enabling the model to capture complex relationships between personality traits and sentimental expressions. It also prevents the model from deviating from the true sentiment category due to overfitting on personality traits. Removing this constraint leads to excessive focus on aligning personality with sentiment, causing an imbalance between matching personality features and accurately classifying sentiments, resulting in decreased performance.

\begin{figure}[t]
\centering
\includegraphics[width=3.3in, keepaspectratio]{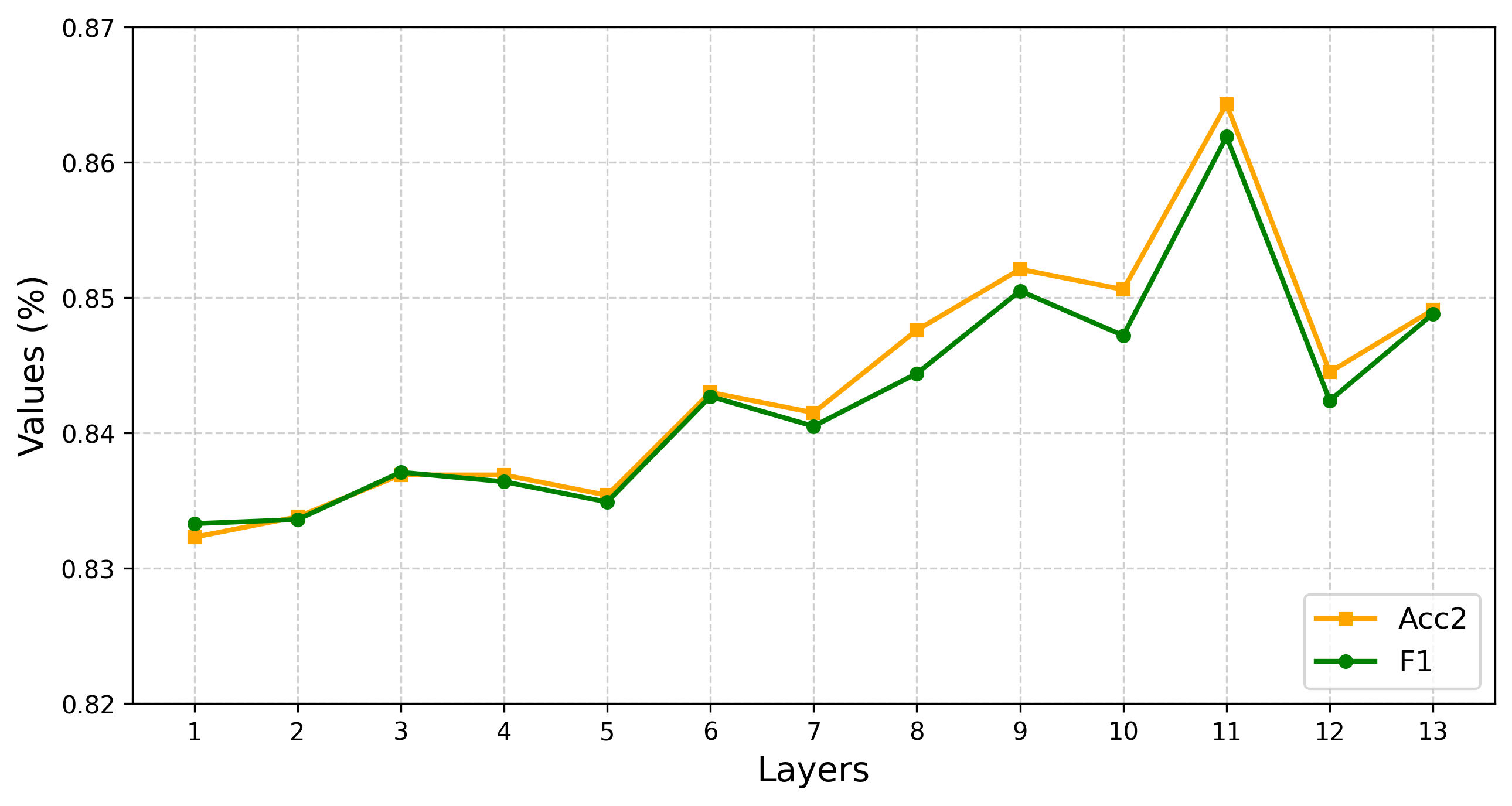}\\
\caption{Acc2 and F1 for personality-sentiment alignment applied at different layers on the CMU-MOSI.}
\label{fig3}
\end{figure}

Removing the multimodal contrastive learning loss decreased performance, with Acc2 dropping by 1.39\% and F1 by 1.27\%. This suggests contrastive learning during multimodal pre-fusion helps align different modalities. By reducing unimodal noise and enhancing complementary sentimental features, the model establishes semantic relationships between modalities during later cross-modal interactions.

\subsection{Layer Analysis}
This study investigates the effects of the personality-sentiment alignment strategy at different network layers through a hierarchical experiment. Figure \ref{fig3} shows the changes in Acc2 and F1 metrics for each layer's alignment strategy on the test set. The model architecture consists of 13 layers, with alignment strategies as follows:

Layers 1-11: Personality embeddings are aligned with text embeddings outputted by the text encoder. The alignment effect in the shallow layers is relatively weak, as the shallow features of the text encoder have not captured sufficient sentimental semantic information.

In particular, in pure text processing layers (Layers 1-11), as network depth increases, model performance shows a stable upward trend, peaking at Layer 11. This confirms that deeper text encoders can extract richer sentimental semantic features, enhancing personality-sentiment alignment.

Layer 12: Personality embeddings align with multimodal features obtained from concatenated visual and audio features. Performance in this layer significantly declines, likely due to modality noise in unfused visual and audio features, making cross-modal semantic matching difficult.

Layer 13: Personality embeddings align with fused features from multimodal encoding. Performance in this layer improves compared to Layer 12 but remains below Layer 11. Analysis attributes this to personality embeddings serving as pure text representations, creating semantic space discrepancy with multimodal joint features.

\section{Conclusions}
This paper presents PSA-MF, an MSA method that incorporates personality information and employs a multi-level fusion strategy to achieve personalized sentimental feature extraction. It extracts personality traits from text and maps personality embeddings into the sentimental representation space through a personality-sentiment alignment module. The multi-level fusion strategy enhances the flow of personality information across modalities and facilitates the progressive fusion of personalized sentimental information. Through comparative experiments and ablation studies, we validate the superiority of this method. Our method provides a detailed approach and comprehensive fusion strategy to understand and analyze multimodal sentiments. Future work could extend personalized sentiment analysis to other modalities and explore new fusion strategies.

\bibliography{main}

\end{document}